\documentclass{iaus}
\usepackage{graphicx}
\def\farcs{\hbox{$.\!\!^{\prime\prime}$}}
\def\arcsec{\hbox{$^{\prime\prime}$}}
\def\degree{\ensuremath{^\circ}\/}

\title[Magnetic structure of prominence threads] 
{A first look into the magnetic field configuration of prominence threads using spectropolarimetric data}

\author[D. Orozco Su\'arez]   
{D. Orozco Su\'arez$^{1,2}$, A. Asensio Ramos$^{1,2}$, \and J. Trujillo Bueno$^{1,2,3}$}
 
  \affiliation{$^1$Instituto de Astrof\'isica de Canarias,   E-38205 La Laguna, Tenerife, Spain, \\[\affilskip]
   $^2$Departamento de Astrof\'isica, Universidad de La Laguna, 
  E-38206 La Laguna, Tenerife, Spain \\[\affilskip]
  $^3$ Consejo Superior de Investigaciones Cient\'ificas, Spain}

\pubyear{2013}
\volume{300}  
\pagerange{1--4}
\jname{Nature of Prominences and their Role in Space Weather}
\editors{B. Schmieder, J.-M. Malherbe \& S. Wu, eds.}
\begin{document}

\maketitle

\begin{abstract}
We show preliminary results of an ongoing investigation aimed at determining the configuration of the magnetic field vector in the threads of a quiescent hedgerow solar prominence using high-spatial resolution spectropolarimetric observations taken in the He~I~1083.0~nm multiplet. The data consist of a two-dimensional map of a quiescent hedgerow prominence showing vertical threads. The observations were obtained with the Tenerife Infrared Polarimeter attached to the German Vacuum Tower Telescope at the Observatorio del Teide (Spain). The He~I~1083.0~nm Stokes signals are interpreted with an inversion code, which takes into account the key physical processes that generate and/or modify circular and linear polarization signals in the He~I~1083.0~nm triplet: the Zeeman effect, anisotropic radiation pumping, and the Hanle effect. We present initial results of the inversions, i.e, the strength and orientation of the magnetic field vector along the prominence and in prominence threads. 
\end{abstract}

\firstsection 
\section{Introduction}

Quiescent hedgerow prominences are sheets of plasma characterized by the presence of thin, vertically oriented threads  that show them as highly dynamic structures (e.g., \cite[Berger et al.\ 2008]{2008ApJ...676L..89B}, \cite[Chae et al. 2008]{2008ApJ...689L..73C}, and \cite[Berger et al.\ 2010]{2010ApJ...716.1288B}). However, the observed threads evolution is at odds with what current theoretical prominence models predict: magneto-static structures where, as for instance in the classical \cite[Kippenhahn \& Schl\"uter (1957)]{1957ZA.....43...36K} and \cite[Kuperus \& Raadu (1974)]{1974A&A....31..189K} models, the magnetic pressure exerted by horizontal, bowed field lines acts as the plasma supporting mechanism (for reviews on prominence models see \cite[Mackay et al. 2010]{2010SSRv..151..333M} and \cite[Labrosse et al.\ 2010]{2010SSRv..151..243L}). Several authors have provided plausible physical mechanisms for the transport of plasma through prominence sheets. For instance, the Rayleigh-Taylor instability explains how hot plasma (plumes) and magnetic flux can be transported upwards from prominence bubbles (Hillier et al.\ 2012a, 2012b) or how the condensation of hot coronal plasma into the prominence can produce falling plasma droplets (\cite[Haerendel \& Berger 2011]{2011ApJ...731...82H}).

If the highly dynamic plasma is coupled to the prominence magnetic field, we somehow expect the magnetic field to show local variations, at scales comparable to or smaller than the typical sizes of prominence threads. However, prominence fine scale structures (such as threads) have been elusive to magnetic field measurements mainly because of the low spatial resolution achieved in full spectropolarimetric ground-based observations.  Thus, we have information about the global magnetic structuring only (Tandberg-Hanssen \& Anzer 1970, Leroy et al.\ 1983, Bommier et al.\ 1994, Casini et al.\ 2003, Athay et al.\ 1983, Leroy et al.\ 1983, Casini et al.\ 2003, 2005, Merenda et al.\ 2006, 2007, Orozco Su\'arez et al.\ 2013). Knowing the magnetic field configuration in prominence threads is important in order to test the different proposed physical scenarios for the threads dynamic behaviour.  Here, we present preliminary results of an ongoing investigation aimed at determining the configuration of the magnetic field vector in the threads of a quiescent hedgerow  prominence. 

\section{Observations and data analysis strategy}

To infer the magnetic field configuration in prominence threads we used spectropolarimetric observations taken in the 1083 nm spectral range. This spectral range contains the He I 1083.0 nm triplet which is sensitive to the joint action of atomic level polarization (i.e., population imbalances and quantum coherences among the levelÕs sublevels, generated by anisotropic radiation pumping) and the Hanle (modification of the atomic level polarization due to the presence of a magnetic field) and Zeeman effects. In this triplet, the Stokes Q and U signals detected in prominences are dominated by atomic level polarization and the Hanle effect while Stokes V is mostly dominated by the longitudinal Zeeman effect. This makes the He I 1083.0 nm triplet to be sensitive to a wide range of field strengths, from dG to kG. Moreover, we know the physics of the Hanle and Zeeman effects in this triplet, which is described in detail in Trujillo Bueno et al.\ (2002), Socas-Navarro et al.\ (2004), Trujillo Bueno \& Asensio Ramos (2007), and in the book of quantum theory of polarization by Landi and Landolfi (2004). The He~I~1083.0~nm triplet has already provided remarkable results regarding the magnetic field structure of quiescent and active region filaments and prominences (e.g.,  Lin et al.\ 1998; Trujillo Bueno et al.\ 2002; Merenda et al.\ 2006, Orozco Suarez et al.\ 2012, 2013, Kuckein et al.\ 2009, 2012a,b, Xu, Z. et al.\ 2012). 

\begin{figure}[!ttp]
\vspace*{0.5 cm}
\begin{center}
\includegraphics[width=13cm]{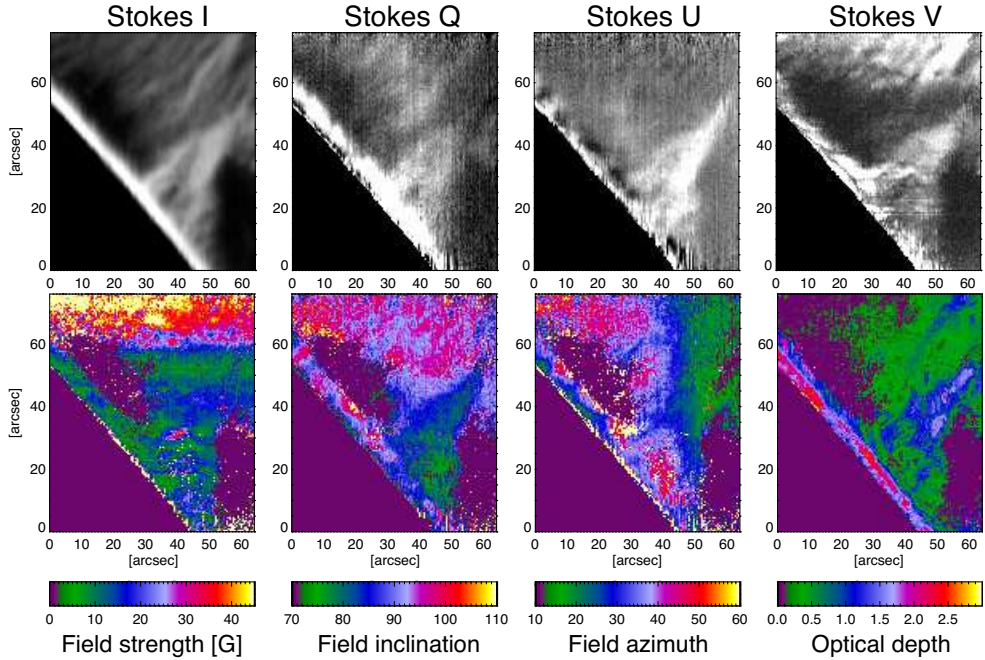} 
\caption{Top panels display the Stokes I peak intensity and the integrated Stokes Q, U, and V absolute value signals corresponding to the observed quiescent prominence. Bottom panels show the inferred field strength, inclination and azimuth values corresponding to the prominence as well as the optical depth.  Black (purple) colour in the top (bottom) panels represents the solar disk as well as pixels with little signal (and therefore excluded from the analysis). Note that the inclination is measured with respect to the local vertical and the azimuth is measured counterclockwise with respect to the line of sight direction.}
\label{fig1}
\end{center}
\end{figure}

The data we present here correspond to a two-dimensional map of a quiescent hedgerow prominence. They were taken at the German Vacuum Tower Telescope at the Observatorio del Teide (Spain) with the Tenerife Infrared Polarimeter (Collados et al.\ 2007). The pixel sampling is 0.5 arcseconds and the spectral sampling 1.1 picometers. The exposure time per slit position was about 30 seconds that allowed us to achieve high signal-to-the-noise ratios in linear and circular polarization. It took 1.5 hours to scan the whole prominence during stable seeing conditions. Figure 1 (top left panel) shows the Stokes I peak intensity corresponding to the 2D map. The field of view is about 60\arcsec$\times$80\arcsec. We estimate a spatial resolution of about 1\arcsec-1\farcs5, which is rather at the limit to detect prominence fine scale structures.  In the map, vertical threads, a prominence foot, and a cavity are clearly distinguishable. The integrated Stokes Q, U, and V absolute value signals are also represented in Fig.\ 1. The Stokes Q and U signals are prominent at the prominence foot (right part). In the prominence body we mostly detect Stokes Q signals. Interestingly, there is a remarkable amount of Stokes V signals in the upper part of the prominence. Note that, close to the solar limb, there is a considerable amount of Stokes signals corresponding to solar spicules. 

For modelling and interpreting the He I 1083.0 nm triplet polarization signals we have applied an user-friendly diagnostic tool called "HAZEL" (from HAnle and ZEeman Light, Asensio Ramos et al.\ 2008). This code allows us to determine the strength, inclination and azimuth of the magnetic field vector in all pixels with measurable linear polarization signals. It takes into account the key physical processes that generate and/or modify circular and linear polarization signals in the He I 1083.0nm triplet: the Zeeman effect, anisotropic radiation pumping, and the Hanle effect. The radiative transfer is taken into account assuming a suitable slab model.

\section{Preliminary results and future work}

Figure 1 (bottom panels) shows the strength, inclination, and azimuth of the field vector inferred with the HAZEL inversion code. We display only the horizontal solution\footnote{When interpreting the He I 1083.0 nm multiplet, there are several magnetic field vector configurations compatible with the observed polarization signals. These result from the 180\degree\/ azimuth ambiguity and for the so-called 90\degree\/ ambiguity of the Hanle effect. The latter provides two solutions which differ by about 90 degree in the field inclination, leaving us with a  horizontal and vertical solutions. A detailed explanation about the existing ambiguities in the He I 1083.0 nm triplet can be found in Asensio Ramos et al.\ (2008).}  since it is in line with most model predictions, i.e., with the fact that the prominence material should be sitting in dipped magnetic field lines. The field strength is about 10-15 Gauss and increases up to 40 Gauss in the upper part of the prominence. The field vector is about 90 degree inclined with respect to the local vertical in the prominence body. It becomes slightly more vertical at the foot. Finally, the field vector azimuth is about 30 degree. This result roughly agrees with previous measurements (e.g., Casini et al. 2003). In the same figure we show the optical depth of the He I 1083.0 nm triplet. We obtain optical depths of the order of 0.5 in the prominence body and of about 1.5 in the prominence foot. 
 
 In Figure 2 we show a zoom corresponding to a central part of the prominence. Note that the image has been rotated in order to show the threads vertically aligned. At least, two threads can be outlined in the peak intensity plot. However, at a glance we do not appreciate neither local variations in the field vector or correlations between the thread pattern in the intensity images and the magnetic field vector. Interestingly, there is a field strength gradient (the field strength decreases from left to right). This gradient in the field strength is uncorrelated with the peak intensity or with the inclination and azimuth maps. The inclination of the field is about constant. The azimuth also shows a slight variation from the left (30\degree--40\degree) to the right (15\degree--20\degree). In summary, we can conclude that the magnetic field along the prominence body where there are vertical threads is rather homogeneous. This suggests that the plasma concentrations that give rise to the threads are detached from the magnetic field. This would be in line with the magneto-convection scenario proposed by Berger et al.\ (2008). Since this is the first measurement of the field vector in quiescent hedgerow threads we wonder whether long integration times and/or lack of spatial resolution may be hampering the analysis. For instance, highly dynamic structures (such as threads) may washout the magnetic field vector information encoded in the measured Stokes parameters. This possibility seems not very plausible because we spatially resolve the prominence threads. These data are being analysed in very much details and we hope to publish the final results very soon.

As a concluding comment, we highlight that presently it is possible to determine the strength and orientation of the magnetic field in solar prominences by interpreting the Stokes I, Q, U, and V profiles of the He I 1083 nm triplet. In this case, we have analyzed data corresponding to a quiescent hedgerow prominence. The data allowed us to look at the magnetic configuration of prominence threads. We found that the orientation of the magnetic field vector is constant and does not show small-scale spatial variations at a resolution of 1\arcsec--1\farcs5.  Certainly, full spectropolarimetric observations in the He~I~1083.0~nm multiplet at very high spatial resolutions such as those achievable with GREGOR at present time or EST, ATST, and SOLAR-C in the near future, will help up to determine the local variations (if any) of the magnetic field vector in solar prominences.

\begin{figure}[!ttp]
\vspace*{0.5 cm}
\begin{center}
\includegraphics[width=13cm]{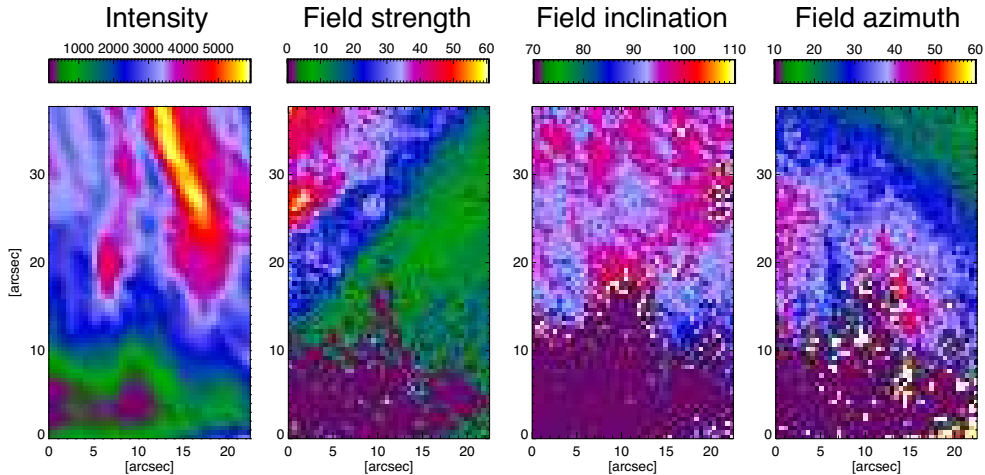} 
\caption{Intensity map showing a zoom to vertical prominence threads and the inferred magnetic field strength, inclination, and azimuth. Note that the map has been rotated to put the threads vertically. In the field strength, inclination, and azimuth maps the purple areas correspond to pixels with no emission in the He I 1083.0 nm triplet.}
\label{fig1}
\end{center}
\end{figure}


\end{document}